\documentstyle[prl,aps,epsf]{revtex}
\begin{document}

\twocolumn[\hsize\textwidth\columnwidth\hsize\csname@twocolumnfalse%
\endcsname

\title{
Current Induced Magnetization Switching in Small
Domains of Different Anisotropies
}

\author{Ya.B.Bazaliy, B.A.Jones, and Shou-Cheng Zhang}

\address{Department of Physics, Stanford University, Stanford, CA 94305\\
  IBM Almaden Research Center, 650 Harry Rd. San Jose, CA 95120}
\date{\today}
\maketitle
\begin{abstract}

  Several recent experimental studies have confirmed the possibility
  of switching the magnetization direction in the small magnetic
  domains by pumping large spin-polarized currents through them.  On
  the basis of equations proposed by J.Slonczewski for the case when
  magnetization of the domains is almost uniform, we analyze the
  stability and switching for two types of magnetic and shape
  anisotropies of a magnetic domain in a nanowire and find
  qualitatively different behavior, including different shapes of
  bistable regions. Our study is analytic as opposed to recent numeric
  work. Assumed anisotropies can be realized in experiments and our
  predictions can be used to experimentally test the theory of
  spin-transfer torques. Such test would be especially interesting
  since alternative approaches are discussed in the literature.

\end{abstract}

]

Recently considerable experimental interest
\cite{tsoi,wegrowe,ralph,sun} has been shown in the torques created by
spin-polarized currents in a magnet.  This interest is fueled in part
by the proposals of developing a convenient writing process for
advanced metallic magnetic RAM \cite{slon-patent} where the reading
process will be based on the magnetoresitance or other effect
\cite{GMR-heads}. A general theoretical framework for the description
of such ``spin-transfer'' torques is set in \cite{slon,berger,bjz}.

One of the particular experimental setups in which this effect can be
studied is a thin ($ \le 300 $ nm) normal metal wire with two magnetic
pieces embedded in it (see Fig.1). If the distance between the
magnetic pieces does not exceed the spin diffusion length in the
normal spacer between them, a current passing through the wire will
induce spin-transfer torques in both magnets. Such setup was
originally considered in \cite{slon} for the case when both magnetic
pieces are isotropic and their magnetizations are initially not
collinear. It was predicted that both magnetizations will perform
rotation in a fixed plane keeping the angle between them constant.
\begin{figure}[h]
  \vspace{-0.3cm} \hspace{0.5cm} \epsfxsize=7cm \epsfbox{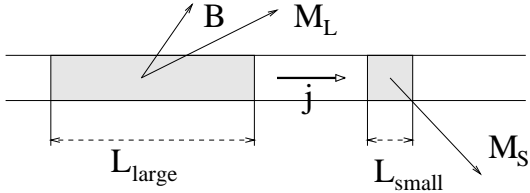}
  \vspace{0.1cm}
\caption{Model experimental setup. Current $j$ is passed through the
  nanowire with two magnetic pieces (shaded areas) with magnetizations
  ${\bf M}_L$ and ${\bf M}_S$. External magnetic field $\bf B$ can be
  applied in an arbitrary direction.}
\end{figure}
However for a real material one must also take into account
magnetocrystalline and shape anisotropy. In this paper we present
the first exact solutions with anisotropies taken into account. Attempts
to incorporate them were made in the experimental reports
\cite{ralph,sun} and an extensive treatment of a particular
experimental situation was given in a numerical study
\cite{sun-theor}.

If the size of the pieces is larger then the domain wall width, the
magnetization may be not uniform throughout the piece.  In this case
equations from \cite{bjz} must be used inside each piece to determine
the magnetic configuration.  In this work we will assume that magnetic
pieces are sufficiently small and treat them as magnetically uniform.
Then the total anisotropy, including magnetocrystalline and shape
contributions, will be given by a symmetric tensor $K^{(tot)}_{ik}$.

The problem simplifies if one of the magnetic pieces is much longer
then the other. At a given current density $j$ the spin-transfer
torque $T_{st}$ is proportional to the cross-section of the piece
while the torques $T_a$ due to anisotropy are proportional to its
volume.  The ratio $T_{st}/T_{a} \sim 1/L$, where $L$ is the
length of the piece, thus the small piece will be affected by the
spin-transfer torque starting from a much smaller value of $j$. We can
therefore neglect the effect of $T_{st}$ on the large piece, called a
polarizer, and assume its magnetization to be constant. Magnetizations
of the large and small pieces will be denoted as ${\bf M}_L$ and ${\bf
  M}_S$ respectively.

To test the theory, one would like to be able to control the direction
of the magnetization of the polarizer with respect to the anisotropy
directions of the small piece. The easiest way to change ${\bf M}_L$
is an application of an external magnetic field $\bf B$. Of course,
$\bf B$ will also act on the small piece and must be taken into
account in the equations of motion. The properties of a system with a
given anisotropy $K_{ik}$ can ultimately be presented as a phase
diagram in the $(j,{\bf B})$ space with spin-transfer effects
determined by the magnitude of the current and by ${\bf M}_L = {\bf
  M}_L({\bf B})$. In this paper we calculate a section of such diagram
for certain directions of $\bf B$ and anisotropies. Note, that for
technical applications in the memory writing process one is interested
in finding anisotropy tensors which satisfy two conditions: (a) there
is a section of the phase diagram at a fixed external field ${\bf B} =
{\bf B}_0$ where ${\bf M}_S$ is bistable at $j = 0$; (b) by passing a
current $j \neq 0$ one can switch between these two metastable states.
However for the purposes of testing the theory it is better to start
with the cases where the phase diagram can be calculated exactly and
compare theoretical and experimental results.

In \cite{sun-theor} stability of the zero-current equilibrium position
of ${\bf M}_S$ was studied by an approximate analytical method and by
numerical solution of a differential equation for the time evolution
of ${\bf M}_S$. The latter gave the final position of magnetization
and the details of the switching process. Due to the approximation
used, there was a discrepancy between the switching thresholds
predicted by two methods. In our study we find the equilibrium
positions of ${\bf M}_S$ for each $(j,{\bf B})$ point and analyze
their stability exactly.  Knowing the nature of all equilibria we are
able to construct the topology of the time-evolution flow of ${\bf
  M}_S$ and predict qualitatively the overall behavior of the system,
including existence of the stable cycles.  Such stable cycles were
predicted in \cite{slon-axial} and observed numerically in
\cite{sun-theor}. Due to energy dissipation they would be impossible
without the current.  However for $j \neq 0$ there is a constant
supply of energy which feeds the periodic motion of ${\bf M}_S$.

{\bf Dynamic Equation for the Small Piece} The magnetic energy of the
small piece is a sum of the intrinsic anisotropy term, shape anisotropy
term, interaction with external magnetic field and exchange
interaction with the large piece.  Approximating the shape of a
monodomain small piece by an ellipsoid we can write
\begin{eqnarray}
\nonumber
\frac{F}{V} &=& \frac{1}{2}\left(-K^{(intr)}_{ik} n_i n_k +
          4\pi M_i N_{ik} M_k \right) -
%\\
%& &
B_i M_i - J_{ex} s_{i} n_i
\end{eqnarray}
where $N_{ik}$ is the demagnetization tensor \cite{LLv8}, $M_i =
M_{Si}$ - magnetization of the small piece, $K^{(intr)}$ -
magnetocrystalline anisotropy tensor, and $J_{ex}$ is the exchange
coupling between the pieces. Vectors ${\bf n}$ and ${\bf s}$ are unit
vectors along the magnetization of the small and large pieces
respectively.  According to \cite{slon}, the modified Landau-Lifshitz
equation for ${\bf M}$ has the form:
\begin{eqnarray}
\label{eq_LLM}
{\dot {\bf M}} &=& -\frac{\gamma}{V}{\dot {\bf L}} =
     \gamma [-\frac{\delta F}{\delta {\bf M}} , {\bf M}] +
\\
\nonumber
&+ &
     \frac{\gamma}{V}\frac{\hbar}{2} A \frac{j}{e} g(P, {\bf s \cdot n})
     [{\bf n} , [{\bf s} , {\bf n}]] +
\tilde\alpha [{\bf n} , \dot{\bf n}]
\end{eqnarray}
where $\gamma = g \mu_B /\hbar$ is the gyromagnetic ratio, $V$ and $A$
are the volume and cross-section area of the piece, the last term
represents Gilbert damping, $P$ is the degree of spin-polarization of
the electrons coming out of the large piece and $g(P,{\bf s \cdot n})$ is
the function derived in \cite{slon}
\begin{eqnarray}
\label{eq_gfunc}
g(P,{\bf s \cdot n}) &=&  \frac{1}{f(P)(3+{\bf s \cdot n}) - 4},
\;\;\;\;
f(P) = \frac{(1+P)^3}{4 P^{3/2}}
\end{eqnarray}
Equation (\ref{eq_LLM}) can be rewritten in terms of $\bf n$
\begin{eqnarray}
\label{eq_LLn}
\dot n_i &=& [{\bf h}, {\bf n}]_i +
           K_{ik} n_k +
    I g(P,{\bf s \cdot n})]
           [{\bf n} , [{\bf s} , {\bf n}]]_i
     + \alpha [{\bf n} , \dot {\bf n}]
\end{eqnarray}
with rescaled coefficients
\begin{eqnarray}
\label{eq_rescaledcoeff}
\nonumber
h_i &=& \gamma (B_i + \frac{J_{ex}}{M} s_i),
\;\;\;
K_{ij} = \frac{\gamma}{M} (K^{(intr)}_{ij} - 4\pi M^2 N_{ij})
\\
I &=& \frac{\gamma}{V}\frac{\hbar}{2} A \frac{j}{e} \frac{1}{M},
\;\;\;\;\;\;\;\;\;
\alpha = \frac{\tilde\alpha}{M}
\end{eqnarray}
The behavior of the small piece will be completely determined by the
directions of $\bf s$ and $\bf h$ with respect to the principal axis
of the anisotropy tensor $\hat K$. Dependence ${\bf s} = {\bf s}({\bf
h})$ is given by the properties of the polarizer.  We will assume that
the damping is small and expand the solutions in $\alpha$.

It was found recently \cite{wireCaxis} that cobalt nanowires grow with
intrinsic easy axis perpendicular to the wire for large wire diameters
$d \geq 50 {\rm nm}$ and with easy axis along the wire for smaller
$d$.  As for the shape anisotropy contribution, it will be an easy
axis along the wire if the length of the small piece $L_S >> d$ and an
easy plane perpendicular to the wire in the opposite case $L_S << d$.
In \cite{ralph,sun-theor} a wire with $d \approx$ 100 nm was used.
This complicates the situation because for any $L_S$ all three
principal components of the total anisotropy tensor are different. On
the contrary, for $d < 50$nm wires anisotropy is always uniaxial. For
$L_S >> d$ it is an easy axis along the wire. For $L_S << d$ the total
constant is given by $K = K^{(intr)} - 4\pi M^2$. If $M$ is
sufficiently large to ensure $K < 0$, one has an easy plane
anisotropy. This is the case for cobalt $K = 5 \cdot 10^6 {\rm
  erg/sm}^3$ and $M = 1.4 \cdot 10^3 {\rm emu}$.

We are going to consider two experimental situations with thin wires.

{\bf Axial Case.} Assume that the polarizer is characterized by an
easy axis anisotropy along the wire. The small piece has uniaxial
anisotropy with respect to the same axis, but the anisotropy constant
$K$ can have either sign. Next, assume that the external magnetic
field is also directed along the wire which itself is oriented along
the $z$-direction. Such situation with $K > 0$ was considered before
in \cite{slon-axial} using a different method.

First we rewrite vector equation (\ref{eq_LLn}) in terms of the polar
angles of ${\bf n}$: $\phi$ and $\theta$. That gives a system:
\begin{equation}
\label{eq_inphiteta}
 \left[
        \begin{array}{cc}
                \sin\theta  & -\alpha
                \\
                -\alpha \sin\theta & -1
        \end{array}
 \right]
     \left\{
            \begin{array}{c}
                    \dot\phi
                    \\
                    \dot\theta
            \end{array}
    \right\}  =
    \left\{
      \begin{array}{c}
         v_{\phi}(\phi,\theta) \\ v_{\theta}(\phi,\theta)
      \end{array}
    \right\}
\end{equation}
To find equilibrium positions one must solve
\begin{equation}
  \left\{
     \begin{array}{l}
         v_{\phi} = \sin\theta (h + K\cos\theta) = 0
          \\
         v_{\theta} = I g(P,\cos\theta) \sin\theta = 0
     \end{array}
  \right.
\end{equation}
When $I \neq 0$ the only stable positions of $\bf n$ on the unit sphere are
the North and South poles, independent of the magnitude of current.

To determine the stability of equilibria we linearize the r.h.s. of
(\ref{eq_inphiteta}) in small deviations.  At $\theta = (0,\pi)$ one
has to use local non-singular coordinates $x = \theta \cos\phi, y =
\theta \sin\phi$.  Linearized equations have the form
\begin{equation}
\label{eq_xydot}
  \left\{
    \begin{array}{c}
      \dot{\delta x} \\ \dot{\delta y}
    \end{array}
  \right\}  = {\hat D}
  \left\{
     \begin{array}{c}
       \delta x  \\  \delta y
     \end{array}
  \right\}
\end{equation}
The nature of equilibria is determined by the eigenvalues $\mu$ of the
dynamic matrix $\hat D$. Real $\mu$'s imply a center or a saddle,
while for complex-conjugate eigenvalues the stationary point is a
stable focus for ${\rm Re}\mu < 0$ or an unstable one otherwise.

Without showing the algebra we write down the result. For the
North pole
\begin{equation}
  \label{eq_eigenvalNorth}
  \mu_N = - I g(P,1) - \alpha(h+K) \pm i |h + K - \alpha I g(P,1)|
\end{equation}
and therefore it is a stable equilibrium for
\begin{equation}
  \label{eq_stabilNorth}
     j > - \frac{\alpha (h+K)}{g(P,1)}
\end{equation}

For the South pole we get a stability criteria
\begin{equation}
\label{eq_stabilSouth}
j < -\frac{\alpha (h-K)}{g(P,-1)}
\end{equation}
The regions of stability of the North and Sough poles are shown on
Fig.2. It is important that there is a region on the diagram where
both equilibrium points are unstable. This necessarily means that
there exists a stable cycle, around which $\bf n$ performs a periodic
motion.

\begin{figure}[h]
\epsfxsize=8.5cm
\epsfbox{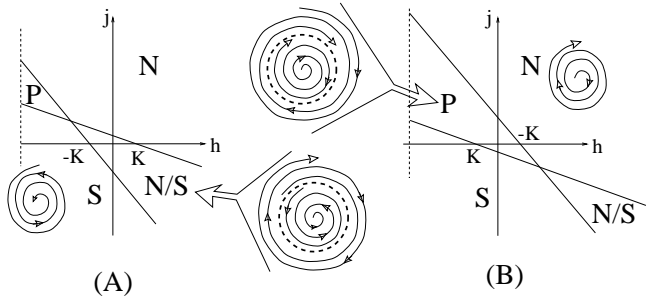}
\vspace{15pt}
\caption{Switching phase diagram for the axial case. (A) easy
axis in the small piece; (B) easy plane in the small piece. The left
boundary (vertical dashed line) represents the switching field of the
polarizer. Regions are marked as: N -- only North pole is stable, S --
only South pole is stable, N/S -- bistable region, P -- precession
region.  Time-evolution flow on the projected sphere (in which N is
mapped to infinity and S to the origin) is shown on the insets.  }
\end{figure}

In general, stability analysis can not give any information about the
shape of the cycle, but in the axial case it is easy to anticipate that it
will be given by $\theta = \theta_0$.  From (\ref{eq_inphiteta}) the
value of $\theta_0$ is given by $I g(P,\cos\theta_0) + \alpha (h + K
\cos\theta_0) = 0$ and the precession frequency by $\omega = \left| I
  g(P,\cos\theta_0)/\alpha \right|$.  When $h$ is decreased through the
precession region at fixed current, $\theta_0$ changes continuously
between $0$ and $\pi$ and $\omega$ increases from $\omega_N = | I
g(P,1)/\alpha |$ to $\omega_S = | I g(P,-1)/\alpha |$. For the $K>0$ case
$I \geq 2\alpha K/(g(P,-1)-g(P,1))$ in the precession region and the
frequency
has a minimal value $2 K g(P,1)/(g(P,-1)-g(P,1))$ of the order of FM
resonance gap. For $K<0$ the frequency can go down to zero since there are
$I=0$ points in the precession region.

{\bf Planar Case.} We now consider a situation when the polarizer is
characterized by an easy-plane anisotropy. The small piece has no
choice but to have $K<0$ too, since by definition $L_S << L_L$ and
shape anisotropy must be even more important for it.  In this
experiment we direct the magnetic field perpendicular to the wire. As a
result, we always have $\bf s$ along $\bf h$ and there is no meaning
to negative values of $h$.  We choose $s$ and $h$ to point in the
$x$-direction.

Functions $v_{\phi}$ and $v_{\theta}$ for the planar case differ
from the axial one. Working out their form we get a system of equations
for the equilibrium points.
$$
%\begin{equation}
  \left\{
     \begin{array}{l}
         v_{\phi} = K\cos\theta\sin\theta - h
         \cos\theta\cos\phi - I g \cdot \sin\phi = 0
          \\
         v_{\theta} = h\sin\phi - I g \cdot
            \cos\theta\cos\phi = 0
     \end{array}
  \right.
%\end{equation}
$$
where $g = g(P,\sin\theta \cos\phi)$. It is that dependence of $g$ on
both $\theta$ and $\phi$ that brings the main complication into the
solution for the planar case.

Both equations are satisfied for $\cos\theta =\sin\phi =0$.
Consequently there is a pair of equilibria: point $A$ $(\phi = 0,
\theta = \pi/2)$ and point $B$ $(\phi = \pi, \theta = \pi/2)$,
positions of which do not depend on the current magnitude.  There are
however additional equilibria given by
\begin{equation}
\label{eq_additnequilibr}
  \left\{
     \begin{array}{l}
         K\sin\theta - \cos\phi (h^2 + I^2 g^2)/h = 0
          \\
        h\sin\phi = I g \cdot \cos\phi\cos\theta
     \end{array}
  \right.
\end{equation}
To deal with the $\phi,\theta$ dependence of $g$ we make a variable
change
\begin{equation}
\label{eq_transform}
  \left\{
     \begin{array}{l}
         \cos\phi\sin\theta = x
          \\
         \sin\phi = y
     \end{array}
  \right.
\end{equation}
Then $g = g(P,x)$. It is an important property of the model and the
planar case geometry, that this substitution succeeds in reducing
system (\ref{eq_additnequilibr}) to a relatively simple decoupled
equation on $x$:
\begin{equation}\label{eq_x}
1 + (\eta x g(P,x))^2 = \frac{x}{x_0}
\end{equation}
with $x_0 = h/K < 0$ and $\eta = I/h$.

Investigation of (\ref{eq_x}) shows that as $\eta$ increases, there
will be one, two or zero roots. We will see below, that switching
happens already at small currents, where only one root exists and
where we can expand the solution as $x = x_0 + \eta x_1 + \ldots$.
This root defines two symmetric equilibrium points $C_{1,2}$ located
as shown on Fig.3 inset. Their positions depend on the magnetic field.
As $h$ grows, both $C$ points move gradually towards $B$ and merge
with it at $h = |K|$.  At higher $\eta$ we observe additional roots -
the full $(I-h)$ diagram will be published elsewhere.

Stability analysis shows, that point $A$ becomes unstable at a
negative current with a magnitude $I_{\rm sw} \sim \alpha$, similar to
the axial case.  That justifies the expansion of $x$ above.  As in the
axial case, we only present the eigenvalues of the dynamic matrix at
all equilibrium points.

For point $A$
\begin{equation}
\mu_A = - I g(P,1) - \frac{\alpha(-K+2h)}{2} \pm
       i \sqrt{h(-K+h)}
\end{equation}
Therefore $A$ is a focus point. Its stability is given by
\begin{equation}
\label{eq_IA}
I > -\alpha\frac{-K + 2h}{2 g(P,1)}
\end{equation}
Linearization near point $B$ gives
\begin{equation}
\mu_B = I g(P,-1) + \alpha\frac{2h + K}{2} \pm \sqrt{-h(h+K)}
\end{equation}
Therefore point $B$ is a saddle point for $h < |K|$ and is a
focus for greater field. In the latter case $B$ is stable if
\begin{equation}
\label{eq_IB}
I < -\alpha\frac{2h + K}{2 g(P,-1)}
\end{equation}
Treatment of both points $C_i$ gives identical results
\begin{eqnarray}
\nonumber
\mu_C &=& \frac{1}{2} \left[ I g\cdot \left(
             \frac{1+\cos^2\theta}{\sin\theta} +
             \sin\theta - f g(P,x_0)\; \cos^2\theta
       \right) -  \right.
\\
\nonumber
 & &   \left. 2 \alpha K + \alpha \frac{h^2}{K}
       \right] \pm i\sqrt{K^2 - h^2 + O(I)}
\end{eqnarray}

In the region of its existence ($h < |K|$) point $C$ is always a
focus. The stability region is given by
\begin{equation}
\label{eq_IC}
I < -\alpha K \frac{x_0(2-x_0^2)}{2+ f g(P,x_0) (1-x_0)^2 x_0}
\end{equation}

\begin{figure}
\vspace{-0.3cm}
\epsfxsize=8.5cm
\epsfbox{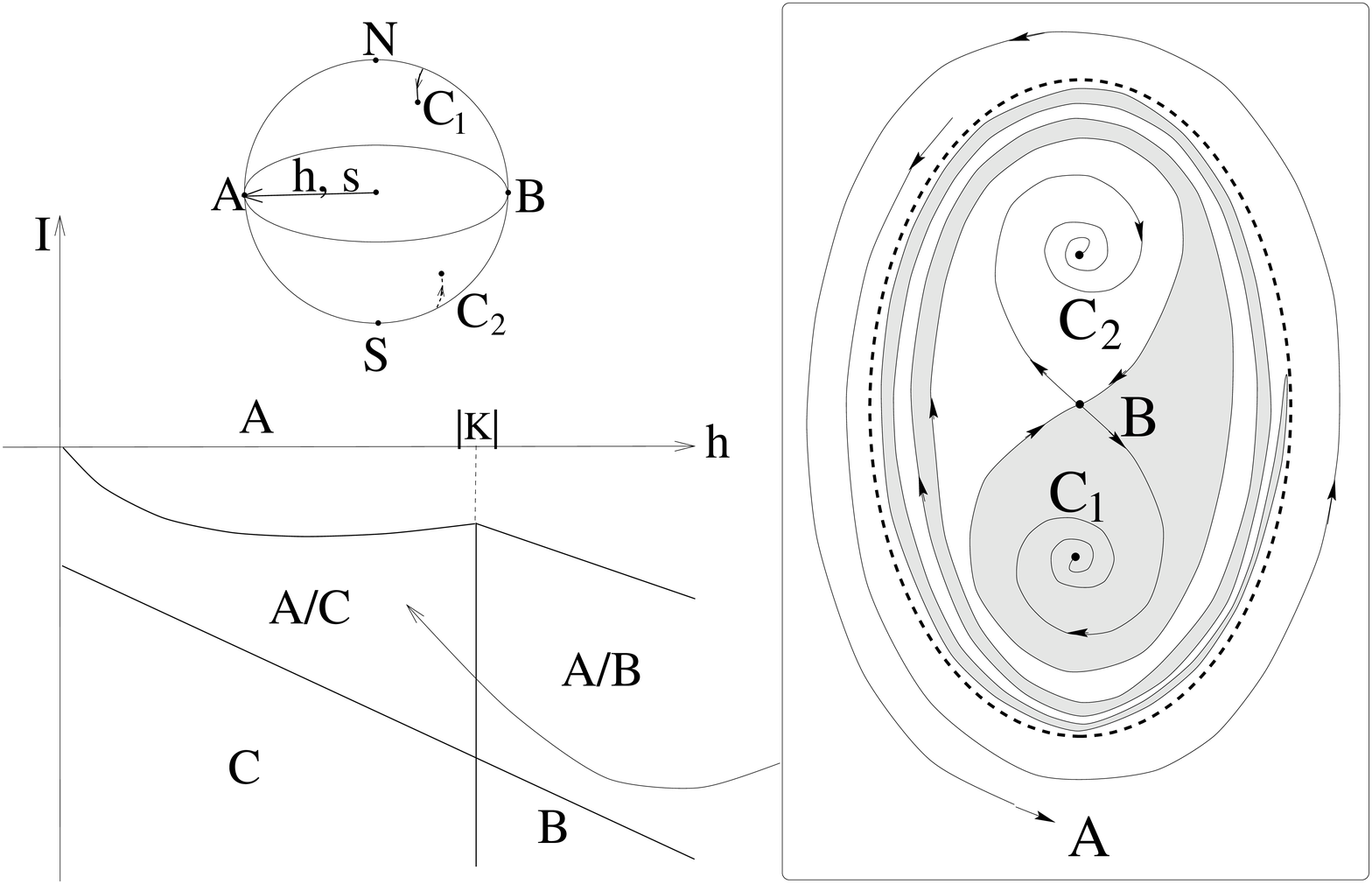}
\vspace{0.7cm}
\caption{
  Switching phase diagrams for the planar case. Regions are marked by
  letters corresponding the stable points. Bistable regions are marked
  by A/C and A/B. Right: time-evolution flow on the projected sphere
  for bistable $A/C$ situation - basin of attraction for $C_1$ is
  shaded.  Upper inset: stable points $A, B, C_{i}(I)$ on the unit sphere.}
\end{figure}

Inequalities (\ref{eq_IA},\ref{eq_IB},\ref{eq_IC}) determine the
switching phase diagram in the small $\alpha$ limit (Fig.3). In the
planar case only bistable regions exist and there is no precession
region on the phase diagram. We also note that the basins of
attraction of $C_1$ and $C_2$ are increasingly interpenetrating near
the boundary separating them from the basin of attraction of $A$
(Fig.3).  In switching from $A$ to $C$ the choice between $C_1$ and
$C_2$ is random.

{\bf Discussion.} Switching patterns depend crucially on the magnetic
anisotropy and the direction of polarization of incoming current. Our
predictions for the axial and planar cases can be used to
experimentally test the theory , in particular the accuracy of the
factor $g(P,{\bf s \cdot n})$ \cite{slon}. This is especially
interesting because alternative descriptions of current driven
excitations are put forth in the literature
\cite{berger-preprint,brauwer}.  Since currently magnetization
directions are not experimentally measured but rather inferred from
the resistive state of the wire, it is important that there are
qualitative differences between axial and planar cases. In the axial
case all boundaries on the $(I-h)$ diagram are straight lines, while in
the planar case one of them is curved. Current sweep through the
precession region in the axial case will show three resistive states
without hysteresis, while a sweep through the bistable region gives
two jumps with a hysteresis. If ${\bf M}_S$ is measured directly,
one will see that it rotates by $180^0$ degrees in the axial case and by
a magnetic field dependent angle $\pi/2 + \theta_0(h)$ in the planar
case. The precession state is a good candidate for observation with a
magnetic force microscope.

The switching current density can be calculated by substituting
({\ref{eq_rescaledcoeff}) into
(\ref{eq_stabilNorth}, \ref{eq_stabilSouth}) and
(\ref{eq_IA},\ref{eq_IB},\ref{eq_IC}). As a
characteristic value one can take
$$
I = \alpha \frac{|K|}{g(P,1)} \;\; \Rightarrow \;\;
  j = \alpha \left( \frac{e}{\hbar} \right)
  \frac{|K^{(intr)} - 4\pi M^2|}{L_S}
$$
For the size of the small piece $L_S = 1 {\rm nm}$, damping $\alpha
= 0.05$ and 40\% polarization degree one gets $j \approx 6.7\cdot 10^7
{\rm A/cm}^2$ using the values of $K^{(intr)}$ and $M$ for cobalt.

It is our pleasure to thank J.Z.Sun, J.Slonczewski and S.P.P.Parkin
for stimulating discussions. Y.B.B. acknowledges support from David
and Lucile Packard Foundation Fellowship in Science and Engineering,
B.A.J. thanks the Aspen Center for Physics for hospitality, S.C.Z. was
supported by NSF grant DMR-9814289.

\end{document}